# Defect Engineered Hexagonal-Boron Nitride Enables Ionic Conduction for Lithium Metal Batteries

Yecun Wu[1†], Yan-Kai Tzeng[2†*], Hao Chen[3†], Kun Xu[4], Gangbin Yan[1], Takashi Taniguchi[5], Kenji Watanabe[5], Arun Majumdar[4,7], Yi Cui[3,6,7*], Steven Chu[1,7,8*]

1. Department of Physics, Stanford University, Stanford, CA, USA
2. SLAC-Stanford Battery Center, SLAC National Accelerator Laboratory, Menlo Park, CA, USA
3. Department of Materials Science and Engineering, Stanford University, Stanford, CA, USA
4. Department of Mechanical Engineering, Stanford University, Stanford, CA, USA
5. Research Center for Electronic and Optical Materials, National Institute for Materials Science, Tsukuba, Japan
6. Stanford Institute for Materials and Energy Sciences, SLAC National Accelerator Laboratory, Menlo Park, CA, USA
7. Department of Energy Science and Engineering, Stanford University, Stanford, CA, USA
8. Department of Molecular & Cellular Physiology, Stanford University, Stanford, CA, USA

†These authors contribute equally to this work.
*Corresponding authors: Yan-Kai Tzeng (ytzeng@slac.stanford.edu), Yi Cui (yicui@stanford.edu), Steven Chu (schu@stanford.edu).

**Abstract**

The practical implementation of lithium-metal anodes has been hindered by uncontrollable dendrite formation and interfacial instability. This study presents a defect-engineering approach of a chemically stable and electrically insulating interfacial layer of hexagonal boron nitride (h-BN) that markedly enhances ionic conductivity through argon ion irradiation. Initially, the electrochemical performance from commercially available, large-area chemical vapor deposition (CVD)-grown h-BN films with industrial-scale argon ion implantation motivated our subsequent detailed investigations using lab-scale exfoliated single-crystal h-BN flakes. Integration of these exfoliated flakes into a hybrid microfluidic-microelectronic chip provided direct evidence that controlled vacancy defects transform h-BN into an efficient lithium-ion conductor while preserving its intrinsic electrical insulation. Experimental validation confirmed improved lithium-metal anode stability, achieving dendrite-free cycling with Li plating/stripping Coulombic efficiencies exceeding 99.5% about

1000 cycles. Further assemble of irradiated h-BN in lithium-sulfur batteries effectively mitigates the polysulfide shuttle effect, sustaining over 97% specific capacity around 300 cycles. These results establish a robust, scalable interface-engineering route for next-generation lithium-metal batteries that combine high ionic transport with excellent electrical insulation.

**Main**

Since the invention of the lithium battery,[1,2] lithium metal anodes have regarded as a highly promising candidate due to their ultrahigh theoretical capacity (3860 mAh g$^{-1}$) and the lowest electrochemical potential among common anode materials.[3] Achieving stable lithium metal anodes could significantly enhance energy density,[4] offering major advantages for next-generation systems such as lithium–sulfur (Li–S) and lithium-oxygen (Li-O) batteries.[5–7]

However, the practical deployment of lithium metal anodes has been consistently hindered by rapid capacity degradation and safety concerns, primarily arising from uncontrolled dendrite growth and interfacial instability.[1] During repeated lithium plating and stripping, lithium tends to form needle-like dendrites capable of piercing the separator, leading to internal short circuits, potential thermal runaway, and cell failure. Moreover, dendritic growth disrupts the solid electrolyte interphase (SEI), which is inherently fragile and dynamic, resulting in rapid capacity fade and low Coulombic efficiency (CE).[8] These challenges have severely limited the viability of rechargeable lithium metal batteries.

As conventional lithium-ion batteries near their theoretical limits, interest in lithium metal anodes has re-emerged over recent decades.[9,10] Ideally, engineered interfacial layers would suppress dendrite formation and stabilize cycling performance. To effectively suppress dendrite growth, a protective interfacial layer must meet several critical criteria.[9] First, it should possess a fracture strength greater than that of lithium metal (~100 MPa);[11] otherwise, it risks being easily penetrated by growing dendrites. Second, the layer must be electrically insulating to prevent lithium from plating on its surface, thereby maintaining its protective function. Third, high ionic conductivity is essential to ensure efficient lithium-ion transport and minimize internal resistance. Lastly, chemical stability within the electrochemical environment is necessary to avoid side reactions that could lead to capacity degradation.

The initial approach involved interconnected amorphous hollow carbon nanospheres deposited onto lithium metal, which were shown to help isolate lithium metal deposits and facilitate the formation of a stable solid electrolyte interphase.[12] Subsequent studies have focused on inorganic materials with strong mechanical properties, such as diamond[13] or lithium composite alloy[14] to effectively suppress dendrite formation.[9,15] Polymer-based materials provide a pliable yet designable artificial SEI that conforms to Li metal and can decouple mechanical strength from fast ion transport through block-copolymer like BAB triblock with A as polyethylene oxide (PEO) and B as polystyrene (PS) or polyelectrolyte (PE) based polymers with lithium salts,[16,17] as well as dynamic-network chemistries.[18,19] Yet, despite decades of extensive research, achieving complete dendrite suppression through interfacial design alone remains elusive, primarily due to trade-offs among ionic conductivity, electrical conductivity, and mechanical and chemical properties. This underscores the need for innovative approaches to overcome these fundamental limitations.

The unique structural characteristics of two-dimensional (2D) materials make them attractive candidates for protective interfacial layers.[20] Among these, hexagonal boron nitride (h-BN) stands out due to its exceptional thermal and chemical stability, along with a high fracture strength of 70.5 GPa,[21] which can reinforce the SEI and mechanically suppress dendrite growth. Unlike electrically conductive 2D materials such as graphene or MXenes, which may trigger Li metal electroplating on their surfaces or undesirable side reactions, h-BN offers excellent electrical insulation, making it a safer and more stable alternative.

While pristine h-BN offers several advantages, it remains intrinsically ionically insulating due to its tightly packed atomic lattice, which lacks mobile species and is nearly impermeable to lithium ions. Prior work has demonstrated that atomic-scale defects in monolayer chemical vapor deposition (CVD)-grown h-BN can serve as conduits for lithium-ion transport.[22] At the same time, native defects such as vacancy clusters and grain boundaries in monolayer h-BN can act as nucleation sites for lithium dendrites. Utilizing multilayer h-BN can help seal these grain boundaries and reduce nucleation, yet maintaining sufficient ionic conductivity across multiple layers remains challenging. As a result, the creation of well-controlled defects that enable efficient Li$^+$ transport through multilayer h-BN is essential for realizing its potential in practical applications.

This work introduces a fundamentally distinct strategy: employing argon (Ar$^+$) ion irradiation to engineer h-BN into an ionically conductive yet electronically insulating interfacial layer. Ion beam irradiation is a well-established technique for doping semiconductors and introducing defects in wide band gap materials, particularly for quantum and optoelectronic applications.[23–25] Building on this principle, h-BN is irradiated with Ar$^+$ ions to generate a network of defects that enable Li$^+$ transport, while the natural interlayer spacing of h-BN functions as a filter to facilitate directional ion conduction, as illustrated in Fig. 1. Importantly, the wide band gap of h-BN is preserved post-irradiation, maintaining its excellent electronic insulation.[26] Consequently, the Ar$^+$-treated h-BN exhibits dual functionality: it promotes Li$^+$ conduction while effectively blocking electron flow, thereby fulfilling the core requirements of a dendrite-suppressing interfacial layer. Therefore, Li metal deposition is expected to take place under the defective h-BN layer which protects Li metal from reacting with organic electrolytes.

Understanding defect formation in h-BN under Ar$^+$ irradiation began with Monte Carlo simulations.[27] Fig. 2a presents the trajectories of one thousand 30 kV Ar$^+$ ions incident normally on a 40 × 40 nm$^2$ h-BN region. As the ions pass through the material, energy loss occurs progressively due to collisions with lattice atoms, illustrated by the color transition of trajectories from red to blue. These collisions lead to the formation of vacancies within the h-BN lattice. After multiple scattering events, most ions diverge from the focused beam path and exit the material.

Simulated vacancy density profiles at varying accelerating voltages, accounting for both direct impacts and damage cascades, are shown in Fig. 2b. At low voltages, Ar$^+$ ions tend to cause surface-level damage and become embedded near the top layer of h-BN. At an intermediate voltage (~30 kV), vacancy generation peaks at a depth of around 20 nm. In this regime, fast-traveling ions pass quickly through the surface region, resulting in a lower vacancy concentration at the entrance, followed by a peak in the mid-layer and a subsequent decline as energy dissipates. At higher voltages, the created vacancy profile appears more uniform, as the ion range exceeds the total thickness of the h-BN film.

The experiments on lab-scale exfoliated h-BN falke employed an Ar$^+$ ion with 30 kV accelerating voltage, the maximum routinely available in research-scale plasma-focused ion beam (PFIB) systems. For battery industry applications, the process was transferred to a commercial ion implantation system capable of reaching up to 100 kV. As shown in Fig. 2b, the difference in vacancies between 30 kV and 100 kV remains within one order of magnitude, supporting the scalability and consistency of the defect-engineering approach. Further characterization of Ar$^+$-induced defects was conducted using scanning transmission electron microscopy (STEM) to directly visualize changes in the atomic structure as shown in Fig. 2c and Extended Data Fig. 1. Vacancies were observed; however, due to the $AA'$ stacking of h-BN and the multilayer nature of the sample, it is challenging to accurately quantify the type and number of defects. Although the defect density varied, the overall defect structure remained relatively unchanged for doses below 10$^{15}$ cm$^{-2}$. At 10$^{16}$ cm$^{-2}$, however, the h-BN became amorphous without lattice and could no longer be clearly resolved in TEM.

Single-crystal exfoliated h-BN flakes were first transferred onto custom-fabricated silicon nitride chips containing through-holes to create free-standing regions suitable for optical characterization. Figure 2d displays photoluminescence (PL) spectra of these flakes following Ar⁺ ion irradiation at 30 kV across a range of doses. In pristine h-BN, a sharp Raman feature appears near 538 nm, corresponding to the E₂g vibrational Raman mode at ~1370 cm⁻¹.[28] Upon irradiation, the introduction of vacancies weakens this E₂g peak, while a shoulder emerges near 1340 cm⁻¹ that may attribute to defects.

As the irradiation dose increases from $10^{13}$ to $10^{14}$ cm⁻², a distinct PL peak around 820 nm becomes evident, attributed to the formation of optically active boron vacancies, which are of particular interest in quantum defect studies.[26,29,30] At doses exceeding $10^{15}$ cm⁻², the intensity of this PL signal declines, indicating the formation of multi-vacancy complexes that lack optical activity. By $10^{16}$ cm⁻², the boron vacancy-related PL peak becomes undetectable, consistent with the transition to non-emissive, heavily disordered structures.

Mechanical characterization of irradiated h-BN was conducted using atomic force microscopy (AFM) indentation.[31] A 30-nm-thick h-BN flake was exfoliated and transferred onto a silicon nitride substrate featuring an array of through-hole with a diameter of ~1 μm. Following patterned Ar⁺ irradiation at different doses as in Extended Data Fig. 1a, a diamond-like AFM tip was used to indent the center of the suspended flake, as illustrated in the top panel of Fig. 2e. The corresponding load–displacement curves were recorded.

At thickness of 30 nm, pristine h-BN flakes remain sufficiently rigid that AFM topography scans do not reveal underlying hole boundaries; the flake shows no appreciable curvature even under irradiation doses up to $10^{15}$ cm⁻². However, at a higher dose of $10^{16}$ cm⁻², significant softening occurs, leading to a visible sagging of ~3 nm in the suspended region, enabling the localization of the center of the hole for force mode experiment. Load–displacement data from three independent holes irradiated at $10^{16}$ cm⁻² are summarized in Fig. 2f, with the breaking loads marked by "×". The relationship between load (*F*) and indentation displacement (*δ*) is described by:[21,31]

$$F = \sigma_0^{2D}(\pi a)\left(\frac{\delta}{a}\right) + E^{2D}(q^3 a)\left(\frac{\delta}{a}\right)^3 \qquad (1)$$

where a is the radius of the hole; $\sigma_0^{2D}$ is the 2D pre-tension of the flake; $E^{2D}$ is the effective Young's modulus; $q = 1/(1.049 - 0.15\nu - 0.16\nu^2)$ is a dimensionless constant related to the Poisson's ratio $\nu$. The 3D breaking strength can be approached by using the point-load continuum estimation:[21,31]

$$\sigma_{3D} \approx \frac{1}{t}\sqrt{\frac{F_{break}E^{2D}}{4\pi r}} \qquad (2)$$

where r is the radius of the AFM tip and t is the thickness of the flake[21,31]. By fitting the load–displacement curves with this model (black lines in Fig. 2f), the breaking strength of h-BN irradiated at a dose of $10^{16}$ cm⁻² is determined to be 1.25 ±0.56 GPa. Although this value is substantially lower than the intrinsic strength of pristine h-BN, ~ 70.5 GPa,[21] it is about 10 time of the bending strength of lithium metal (~ 105 MPa). At lower irradiation doses of $10^{15}$ cm$^{-2}$ or lower, our apparatus was unable to break the irradiated film, indicating that irradiated h-BN retains sufficient mechanical robustness to serve as a physical barrier against lithium dendrite growth.

Electrical and ionic conductivity measurements were carried out using a custom-designed hybrid microelectronic–microfluidic platform, schematically illustrated in Fig. 3a, with photos shown in Fig. 3b. Detailed fabrication procedures are provided in Extended Data Fig.2 and described in the Methods section.

To evaluate ionic conductivity, two isolated aqueous reservoirs with 1M LiTFSI were constructed on opposite sides of the chip, separated by an exfoliated h-BN flake to ensure that ion transport occurred exclusively through the modified h-BN layer. Each reservoir contained a lithium iron phosphate (Li$_x$FePO$_4$, LFP)-coated copper electrode, enabling quantitative analysis by sweeping the voltage across the LFP electrodes and recording the resulting ionic current. A 1 M aqueous LiTFSI solution has a pH of ~ 6.0, so the proton concentration (~1 × 10$^{-6}$ M) is roughly six orders of magnitude lower than the 1 M lithium-ion concentration, confirming that Li$^+$ is the dominant charge carrier in the solution.

As a control, a free-standing silicon nitride window (~ 250 nm thick) irradiated at the highest dose ($10^{16}$ cm$^{-2}$) was tested. The measurements confirmed no detectable ionic leakage through the irradiated silicon nitride film (Extended Data Fig. 3a-c), consistent with the shallow penetration depth of 30 kV Ar$^+$ ions in silicon nitride, which is below 50 nm.

Extended Data Fig. 3d represents the current response from a 10-µm open window without any h-BN flakes. In the absence of h-BN, the LFP-solvent-LFP system behaves as a simple electrochemical cell, with lithium intercalation and water-splitting reactions occurring at the electrode interface depending on the applied voltage. A fast cyclic voltammetry scan can suppress the electrochemical reactions while emphasize the ion/charge transport. The corresponding impedance spectrum at open circuit voltage (Extended Data Fig. 3d) matches the response of a Randles equivalent circuit with a low series resistance and large charge transfer resistance due to the small area of the hole, limited the effective surface area of the LFP electrodes.[32]

An intrinsic, non-irradiated h-BN flake was also tested as a baseline, yielding current responses within the instrumental noise level (Fig. 3c, black curve), thereby confirming its intrinsic ionic insulating nature. For irradiation doses between $10^{13}$ and $10^{15}$ cm$^{-2}$, the current–voltage (I-V) characteristics in Fig. 3c exhibit an initial low-current region, followed by an exponential increase in current, transitioning eventually to linear conduction.

Notably, the current with irradiated h-BN at doses $\leq 10^{15}$ cm$^{-2}$ is more than two orders of magnitude lower than that observed in the open-window setup. In the open-window configuration, current-voltage curves measure the resistivity of transporting Li$^+$ ions out of the LFP electrode, through a the small open hole and depositing the ions onto the other electrode. (see Extended Data Fig. 3d). From this data, we infer that the ion transport through the irradiated h-BN is associated with an additional energy barrier, possibly associated with the (partial) de-solvation process required for Li$^+$ ions to transfer across the electrolyte/h-BN interface. This behavior can be modeled as two diodes connected in a reverse configuration, where ion entry into h-BN incurs an energy penalty, while exit does not, as illustrated in the equivalent circuit in Fig. 3d.[33]

In typical battery operation, this energetic penalty is partially offset by exothermic reduction of partially de-solvated Li$^+$ at the electrode surface, masking the intrinsic de-solvation barrier in voltage–current profile. In contrast to a typical batter configuration where the SEI layer possible other protective film is not in direct contact with the anode, the platform used here decouples ion de-solvation from charge transfer. Since h-BN is electrically insulated, this apparatus enables direct quantification of solvation energy and its impact on ion transport. Once the applied electric field surpasses the energy barrier of as shown in Fig. 3c, the current becomes dominated by the ionic resistance.

To extract ionic conductivity, the forward and reverse I-V branches are averaged to cancel capacitive contributions, and the high-field region (>2.4 V) is fitted with linear functions, as shown by the dash lines in Fig. 3c. The fitted lines converge at a single intersection point, implicating a common solvation-energy barrier across irradiation doses. . At the highest irradiation dose (10$^{16}$ cm$^{-2}$), the I–V characteristics in the inset of Fig. 3c shows reduced current without any diode-like behavior, indicating the removal of the energy barrier previously observed. The voltage scan was limited to 0.25 V to prevent film breakdown caused by excessively high current density in h-BN. This behavior is consistent with TEM and PL characterization, which indicate that h-BN becomes amorphous under such high doses, effectively eliminating the solvation barrier for Li$^+$ ion transport. The resulting ionic resistivities, normalized by flake area and thickness, are summarized in Fig. 3d. With Ar$^+$ irradiation, the h-BN becomes ionically conductive while retaining electronic insulation.

Electrical current-voltage curves for samples irradiated at various doses are presented in Fig. 3e. The data was taken by sweeping the voltage from 0 to 5 V across two photolithographically defined metal contacts, and offset vertically by 100 pA for clarity. In all cases, the measured currents remained at the system noise level (~50 pA).

Based on the results obtained from exfoliated single-crystal h-BN, irradiated h-BN demonstrates strong potential as an interfacial layer for lithium-metal anodes. While single-crystal h-BN presents challenges for large-scale applications, multilayer h-BN films used here were directly grown onto the copper foil by a CVD process. Ion irradiation was performed using an industrial ion implantation system operating at an accelerating voltage of 100 kV, replacing the laboratory-scale 30 kV plasma focused ion beam (PFIB) used in earlier experiments. (See Methods, ion irradiation)  Thus, the methods used in this work should be scalable to industrial production.

Electrochemical performance was evaluated in a half-cell configuration, with lithium metal as the counter electrode and a polymer separator placed between the electrodes as shown in Fig. 4a (see Methods). Cells were cycled at a fixed capacity of 1 mAh with a current density of 1 mA/cm². Control samples consisting of bare copper foil showed rapid capacity decay, with failure occurring after approximately 30 cycles due to dendrite-induced short-circuiting. In contrast, CVD-grown h-BN on copper without irradiation failed to support any cycling, confirming its intrinsic ionic insulating behavior (Extended Data Fig. 4a).

As shown in Fig. 4b and Extended Data Fig. 4c, cycle stability began to improve at an irradiation dose of $10^{12}$ cm$^{-2}$, reaching a maximum of ~1000 stable cycles at $10^{14}$ cm$^{-2}$, attributed to enhanced ionic conductivity. At higher doses ($10^{15}$ and $10^{16}$ cm$^{-2}$), performance slightly declined to ~900 cycles, likely due to excessive defect formation and resulting loss of crystallinity. The columbic efficiency (CE) during the initial cycles was relatively low, due to the initial lithiation of h-BN layer.

To further validate the suppression of lithium-metal dendrites growth, the morphology of plated lithium was examined using scanning electron microscopy (SEM). As shown in Fig. 4c, lithium deposited on bare copper exhibits a porous and uneven surface after several tens of cycles, creating favorable condition for dendrite formation. Copper substrates with irradiated h-BN, however, exhibit dense and uniform lithium deposition (Fig. 4d). A high-magnification image (Fig. 4e) reveals the h-BN layer situated atop the deposited lithium metal, confirming that lithium ions successfully penetrate the irradiated h-BN and plate uniformly beneath it.

Top-view SEM images of plated lithium (Extended Data Fig. 5) show that lithium domains formed on irradiated h-BN-coated copper are over an order of magnitude larger than those on bare copper, suggesting improved plating uniformity. To investigate the structural arrangement, a series of X-ray photoelectron spectroscopy (XPS) measurements were conducted at varying etching depths, focusing on the boron 1*s* orbital signal. Fig. 4f shows a gradual increase in boron 1*s* peak intensity as deeper etching removes surface residues, revealing the h-BN layer above the plated lithium.

As a benchmark for evaluating battery performance, the CE was precisely measured in a half-cell featuring h-BN irradiated with Ar$^+$ ions at a dose of $10^{14}$ cm$^{-2}$, using Aurbach's method (Fig. 4g).[34] With an initial lithium plating/stripping capacity of 5 mAh cm$^{-2}$, the cell exhibited a CE of 98.80%. Subsequent cycling at fractional capacities of the initial value, specifically 10%, 50%, and 80%, for 10 cycles each yielding CE values of 99.82%, 99.75%, and 99.49%, respectively.

Impedance plays a critical role in the cycling performance of lithium-metal anodes. Electrochemical impedance spectroscopy (EIS) was employed to evaluate lithium-ion transport resistance on copper electrodes irradiated with varying doses of Ar$^+$ ions (Fig. 4h). At a dose of $10^{14}$ cm$^{-2}$, the EIS spectra reveal an additional resistance of ~7 Ω, while negligible resistance is observed at the higher doses. At 1 mA current, the potential increment caused by h-BN is only a few millivolts. Although the resistance values obtained from EIS are slightly lower than those derived from conductivity measurements, likely due to differences in sample quality between the single-crystal and CVD samples, the two datasets remain closely aligned. This variation is attributed to the solvation energy barrier and the relatively higher current densities inherent to the microfluidic cell geometry used in conductivity measurements.

A full-cell demonstration was next conducted, focusing on lithium-sulfur (Li-S) batteries, which are regarded as promising next-generation energy storage systems due to their high theoretical energy density (~2600 Wh kg$^{-1}$). However, practical deployment remains limited by major challengesthat include lithium dendrite formation on the Li metal anode and the polysulfide "shuttle" movement onto the anode, both of which compromise cycling stability and capacity retention.[35] Beyond protecting the lithium-metal anode, defect-engineered h-BN appears to mitigate the polysulfide shuttle crossover, and is likely due to the very small interlayer spacing (< 0.3 nm) which will not allow lithium polysulfide ions (0.5 -2 nm) to pass through.[36]

To investigate this capability, an optical cell was fabricated as depicted in Fig. 5a. The device featured two solvent reservoirs: one containing 0.1 M lithium polysulfide in 1,3-dioxolane and 1,2-dimethoxyethane (DOL/DME) electrolyte and the other containing pure DOL/DME electrolyte, separated by an irradiated h-BN membrane. Owing to the distinct Raman peaks of lithium polysulfides at approximately 200, 390, and 450 cm$^{-1}$,[37,38] Raman spectroscopy served as a reliable detection method. After 24 hours, Raman spectra (Figure 5b) confirmed that h-BN irradiated at a dose of $10^{14}$ cm$^{-2}$ effectively suppressed polysulfide diffusion, while the membrane irradiated at $10^{16}$ cm$^{-2}$ permitted leakage due to significant structural damage. An additional diffusion test shown in Extended Data Fig. 6a further confirms that h-BN effectively inhibits polysulfide diffusion. These findings suggest that h-BN irradiated at $10^{14}$

$cm^{-2}$ can simultaneously address both dendrite growth and polysulfide shuttle effect in Li-S batteries.

Finally, to examine electrochemical cycling performance, the cycling stability and CE were examined in prototypical Li-S cells. The anode was prepared by depositing 5 mAh $cm^{-2}$ of lithium metal at a current density of 1.25 mA $cm^{-2}$ onto Cu foil with CVD-grown h-BN irradiated at a dose of $10^{14}$ $cm^{-2}$. The cathode consisted of pure liquid $Li_2S_8$ infused into a carbon nanofiber film, with a sulfur mass loading of 5 mg $cm^{-2}$. Each cell was assembled using 40 µL of DOL/DME electrolyte.

After delivering an initial charge/discharge capacity of 1160 mAh $g^{-1}$ at 0.1 C for 20 cycles, the cell was continuously cycled at 0.1 C with a target capacity of 870 mAh $g^{-1}$ (75% of the initial value). The specific capacity remained at ~840 mAh $g^{-1}$ even after 300 cycles. Additional analysis of the cycle life (Extended Data Fig. 6b) shows that cell failure typically occurred around $300^{th}$ cycle, characterized by open-circuit behavior. The voltage profiles at different cycle indices are shown in Fig. 5d. Post-mortem analysis by dissemble the cell in glovebox shown in the inset of Fig. 5c revealed that the primary cause of failure was electrolyte depletion rather than lithium dendrite formation. Further improvements in electrolyte formulation and cell assembly are expected to enhance long-term cycling stability.

**Conclusion**

In summary, a defect-engineering strategy was developed using $Ar^+$ ion irradiation to transform electrically insulating h-BN into a lithium-ion-conductive interfacial layer while preserving its electronic insulating properties. The optimized irradiation process generated a controlled network of sub-nanometer-scale vacancies, facilitating efficient ionic transport without compromising electronic insulation. Both half-cell and full-cell lithium battery evaluations confirmed notable improvements in cycling stability, dendrite suppression, and inhibition of the polysulfide shuttle effect.

Batteries incorporating defect-engineered h-BN demonstrated stable cycling performance with high CE and extended cycle lifetimes. Since failure mode after prolonged cycling was primarily due to electrolyte depletion rather than dendrite-induced short circuits, further lifetime extension might be achieved with advances in electrolyte engineering[39–41] and imporved cell assembly methods. This work presents a promising approach for stabilizing lithium-metal anodes and provides a foundation for advancing high-performance lithium metal and Li-S battery technologies.

**Methods**

Monte Carlo simulation: Ion irradiation behavior was simulated using the Stopping and Range of Ions in Matter (SRIM) software.[27] Displacement threshold energies were set at 19.36 eV for boron and 23.06 eV for nitrogen, based on reported literature values.[42] The density of h-BN was defined as 2.1 g/cm$^3$. For silicon nitride ($Si_3N_4$), standard material parameters available in SRIM were employed. Simulations incorporating full damage cascades were performed to provide accurate predictions of defect formation and ion penetration depth.

Conductivity measurement: The conductivity measurement chip was fabricated from a 300-μm-thick silicon wafer coated on both sides with approximately 250 nm of silicon nitride deposited via plasma-enhanced chemical vapor deposition (PECVD). Photolithography followed by reactive ion etching (RIE) was applied to the backside to define a ~600 μm-wide silicon window. Subsequent wet etching in 30% KOH selectively removed the underlying silicon, leaving a free-standing silicon nitride membrane. On the front side, a 10-μm diameter through-hole was patterned using photolithography and RIE.

A ~30 nm-thick h-BN flake, exfoliated from high-quality single-crystal material, was deterministically transferred onto the through-hole. Electrical contacts were formed by photolithography and deposition of Au/Ti electrodes (50/3 nm thick), followed by a ~20nm-thick $SiO_2$ passivation layer to minimize electrochemical side reactions. Electrolyte reservoirs were constructed on both sides of the chip using PDMS sidewalls and sealed with glass coverslips. Lithium iron phosphate (LFP)-coated copper foil served as the electrode within the reservoirs. Electrical conductivity was measured using a Keysight B1500A Semiconductor Device Analyzer equipped with a four-probe micromanipulator, while ionic conductivity was evaluated via cyclic voltammetry using a BioLogic SP-50e Potentiostat.

The initial cycles typically show higher ionic current, which stabilizes after a few sweeps (Extended Data Fig. 3f), likely due to the initial Li$^+$ intercalation into the h-BN lattice. Reported ionic resistivities are based on these steady-state measurements.

Ion irradiation: For exfoliated single-crystal h-BN flake, Ar$^+$ irradiation was performed using a Thermo Scientific Hydra Plasma Focused Ion Beam (PFIB) system equipped with an argon ion source and operated at an accelerating voltage of 30 kV. The ion dose was controlled by adjusting the irradiation time and beam current. Pristine CVD-grown h-BN films on Cu-foil (5 × 10 cm$^2$) were obtained from SixCarbon Technology (China). Ion bombardment of the CVD samples was conducted by CuttingEdge Ions, LLC, (USA) using an accelerating voltage of 100 kV.

Electrochemical measurement: All electrochemical cells were assembled in a standard 2032 coin-cell configuration. For evaluation of lithium plating/stripping behavior and cycling

Coulombic efficiency, an electrolyte consisting of 1 M lithium bis(trifluoromethanesulfonyl)imide (LiTFSI) dissolved in a 1:1 (w/w) mixture of 1,3-dioxolane (DOL) and 1,2-dimethoxyethane (DME), supplemented with 5 wt% lithium nitrate additive, was employed. Celgard 2325 (25 μm PP/PE/PP) separators were used with 60 μL electrolyte per cell. EIS measurements were conducted using a BioLogic VMP3 system. The CE of lithium-metal electrodes was evaluated using Aurbach's method.[8,34]

Initially, anode current collectors were conditioned by plating 5 mAh cm$^{-2}$ of lithium at 0.5 mA cm$^{-2}$, followed by stripping at the same current density to 1 V. Subsequently, a lithium reservoir ($Q_{reservoir}$) of 5 mAh cm$^{-2}$ was deposited at 0.5 mA cm$^{-2}$. Nine cycles of controlled stripping/plating ($Q_{cycle}$) were then performed at a defined capacity and current densities, followed by complete lithium stripping ($Q_{strip}$) at 0.5 mA cm$^{-2}$ to 1 V. The average Coulombic efficiency over these ten cycles was calculated as:

$$\text{CE} = \frac{(9 \times Q_{cycle}) + Q_{strip}}{(9 \times Q_{cycle}) + Q_{reservoir}} \qquad (3)$$

For lithium–sulfur (Li–S) battery assembly, lithium metal (5 mAh cm$^{-2}$) was first plated onto Cu foil coated with irradiated h-BN at a current density of 1.25 mA cm$^{-2}$. The cells were then disassembled in an argon-filled glovebox, gently rinsed with DOL/DME electrolyte, and immediately reassembled into Li–S cells. Cathodes were prepared by loading liquid $S_8$ onto carbon nanofiber films with a sulfur mass loading of 5 mg cm$^{-2}$, using 40 μL of DOL/DME electrolyte per cell. Cells were initially cycled at a charge/discharge capacity of 1160 mAh g$^{-1}$ at 0.1 C rate.

Characterization: Raman and photoluminescence measurements were conducted using a HORIBA Scientific LabRAM HR Evolution spectrometer. AFM indentation was performed using a Bruker Dimension Icon equipped with a diamond-like hard tip. The spring constant of the tip was calibrated using Sader's method. The sample was first scanned in tapping mode to manually locate the center of the hole, after which the indentation was conducted using Force-Volume mode. The SEM images were performed on a FEI Helios NanoLab 600i DualBeam SEM/FIB. The cross-section SEM images were acquired by using Ga$^+$ beam to cut the sample first. The STEM-LAADF images were acquired by a double-aberration-corrected (scanning) transmission electron microscope (TEAM 1). Dark-field detectors are used to acquire an atomic-scale image. The accelerated voltage is 300 kV.

**Author contributions**

Y.T., S.C., and Y.W. conceived the idea. Y.W. performed the conductivity measurement. Y.T. prepared the samples for battery experiments. Y.T. and H.C. carried out electrochemical measurements. G.Y. helped the preparation of LFP electrodes. K.X. performed TEM characterization with supervision of A.M.; T.T and K.W. provided the h-BN single crystal. Y.T., Y.C., and S.C. supervised the project. Y.W., Y.T., H.C., and S.C. prepared the manuscript with input from all the other coauthors.

**Data availability**

All data needed to evaluate the conclusions are present in the paper and supplementary information. Source data are available from the corresponding author upon reasonable request.

**Acknowledgement**

We acknowledge the 2D FWP support from U.S. Department of Energy (DOE), Office of Basic Energy Sciences, Division of Materials Sciences and Engineering under contract DE-AC02-76SF00515. Yecun Wu acknowledges the support from Stanford Energy Postdoctoral Fellowship and the Precourt Institute for Energy. Yan-Kai Tzeng acknowledges the support from U.S. Department of Energy of the Battery500 Consortium program and Laboratory Directed Research and Development program at SLAC National Accelerator Laboratory, under contract DE-AC02-76SF00515. Kun Xu, Yi Cui and Arun Majumdar acknowledge cryo-EM support from the US Department of Energy, Office of Basic Energy Sciences, Division of Materials Science and Engineering under contract DE-AC02-76SF00515.Part of this work was performed at nano@stanford RRID:SCR_026695. Work at the Molecular Foundry was supported by the Office of Science, Office of Basic Energy Sciences, of the U.S. Department of Energy under Contract No. DE-AC02-05CH11231. Kenji Watanabe and Takashi Taniguchi acknowledge support from the JSPS KAKENHI (Grant Numbers 21H05233 and 23H02052) and World Premier International Research Center Initiative (WPI), MEXT, Japan.

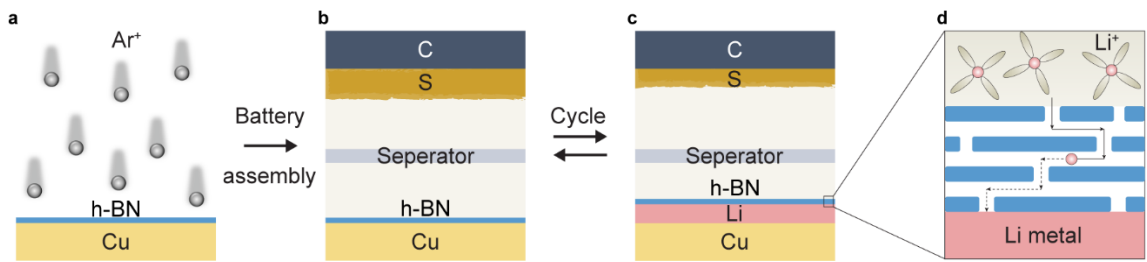

**Figure 1. Schematic illustration of the sample preparation and the proposed mechanism.**
**a,** Ar+ irradiation of h-BN on a copper substrate. **b** and **c,** The irradiated h-BN/Cu is used as the anode in a battery (**b**); after cycling, lithium is plated beneath the h-BN layer (**c**). **d,** The defects (dash circle) in h-BN create pathways for Li$^+$ ion to penetrate the layer, enabling uniform lithium plating and suppressing dendrite formation. The spacing between the planes of h-BN is 0.33 nm. Estimates of the van der Waals gap and the "empty" low-density region (subtracting covalent radii overlap) are ~0.27 nm and ~0.18 nm respectively. This small interlayer spacing is hypothesized to strip away the electrolyte solvation shell but allow isolated Li$^+$ ions to deposit onto the anode.

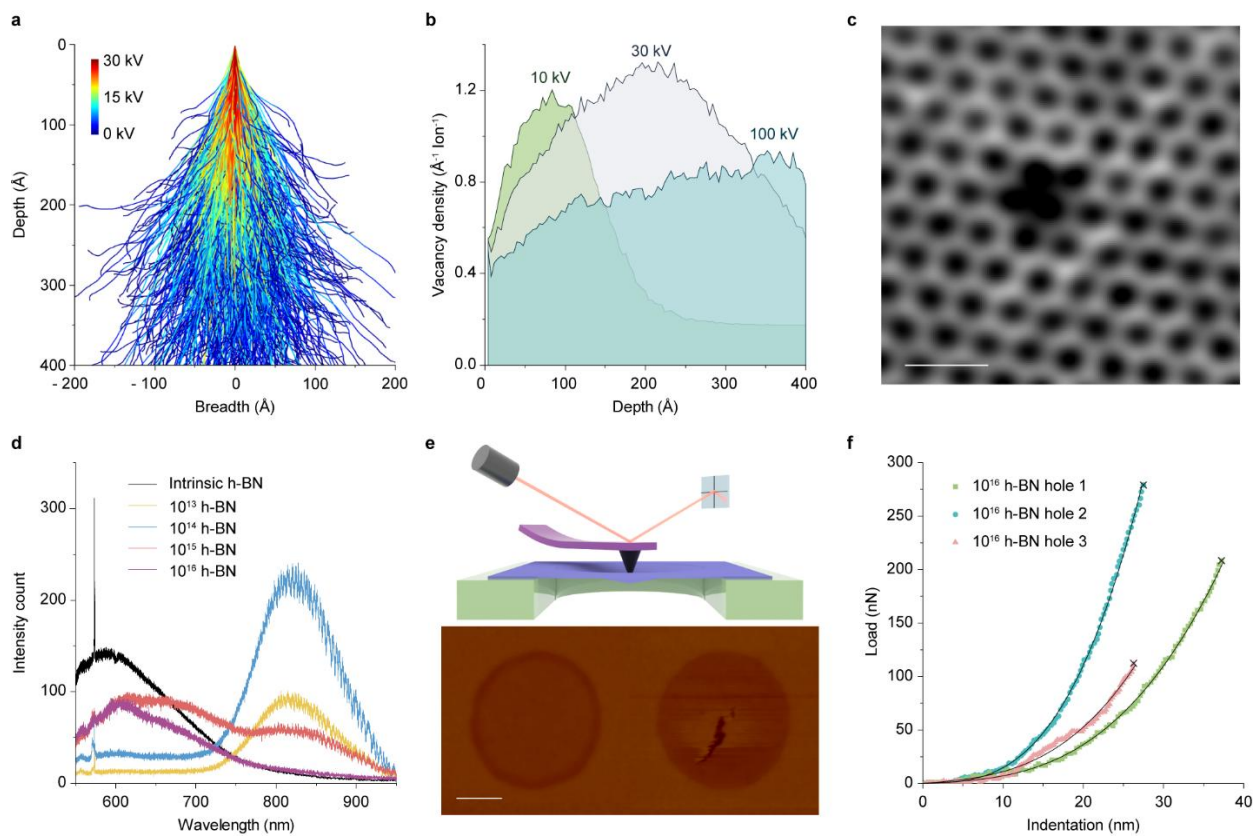

**Figure 2. Defect engineering of h-BN via Ar⁺ ion irradiation.**
**a,** Simulated trajectories of 1,000 Ar⁺ ions (30 kV) incident normal to a 400 × 400 Å h-BN film, with trajectories color-coded by ion energy. **b,** Simulated vacancy-density profiles as a function of depth in the h-BN film, induced by Ar⁺ irradiation at different accelerating voltages. **c,** STEM image of few-layer h-BN after irradiation, showing a three-atom vacancy defect. A bandpass filter was applied to enhance the lattice contrast. Scale bar: 500 pm. **d,** Photoluminescence spectra of free-standing h-BN flakes irradiated with 30 kV Ar⁺ ions at different ion doses. **e,** Schematic illustration of AFM indentation on a free-standing irradiated h-BN film supported by a silicon nitride substrate containing holes. The bottom panel presents AFM images of a pristine h-BN film prior to indentation and a fractured h-BN film after indentation. A film droop of approximately 3 nm is observed at an irradiation dose of $10^{16}$ cm$^{-2}$. Scale bar: 400 nm. **f,** Representative loading curves obtained from three separate holes covered by an irradiated h-BN flake with a dose of $10^{16}$ cm$^{-2}$. The curves exhibit approximately cubic behavior at higher loads, with fracture points marked by ×. The black lines indicate the fitting curves based on Eq. 1.

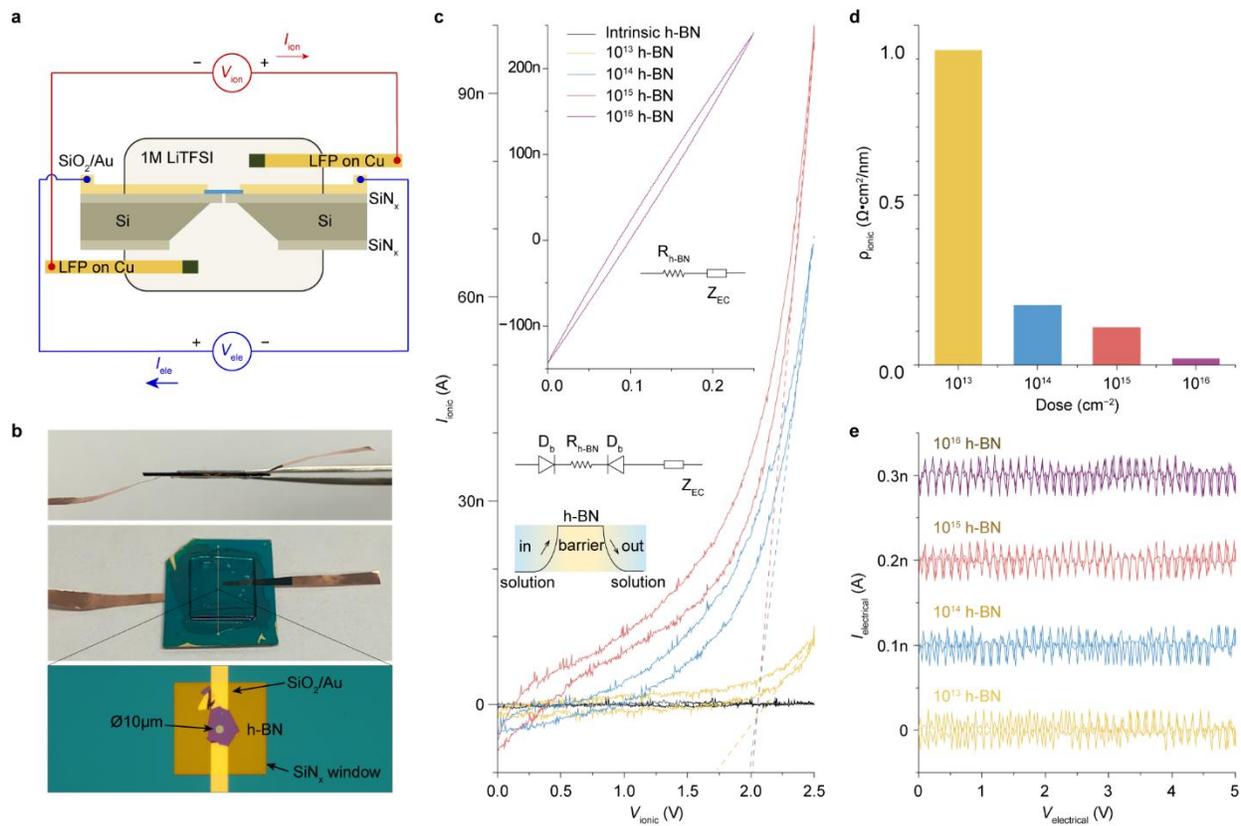

**Figure 3. Conductivity measurements of irradiated h-BN.**
**a,** Schematical illustration of the device setup used to measure the Li$^+$ conductivity through the h-BN film shown in blue. **b,** Photographs showing the side view (top) and top view (middle) of the device, along with a microscope image (bottom) of the free-standing h-BN on a silicon nitride window with gold contacts. **c,** Ionic I-V characteristics of h-BN with different irradiation doses. The data was taken by scanning the voltage from 0 V → 2.5 → 0 volts. The slight hysteresis may be due to the fast scan rate of 100 mV/s (see also extended Fig. 3). The average of the forward and backward scans is used to eliminate hysteresis and determine the ionic conductivity. Dashed lines represent linear fits of the averaged forward and backward sweeps in 2.4~2.5 V range. Inset is the data of h-BN with $10^{16}$ cm$^{-2}$ irradiation dose. The $Z_{EC}$ is the equivalent electrochemical impedance of the open window case in Extended Data Fig. 3e. The $R_{h-BN}$ is the ionic resistivity of h-BN and $D_b$ represents the energy required for Li$^+$ to enter the h-BN layer from solution. **d,** Summary of ionic resistivities for various irradiation doses. Values are normalized by thickness and the area (10 um diameter) of the conductive region. **e,** Electronic I-V characteristics of h-BN at different irradiation doses. Each curve is vertically offset by 100 pA for clarity.

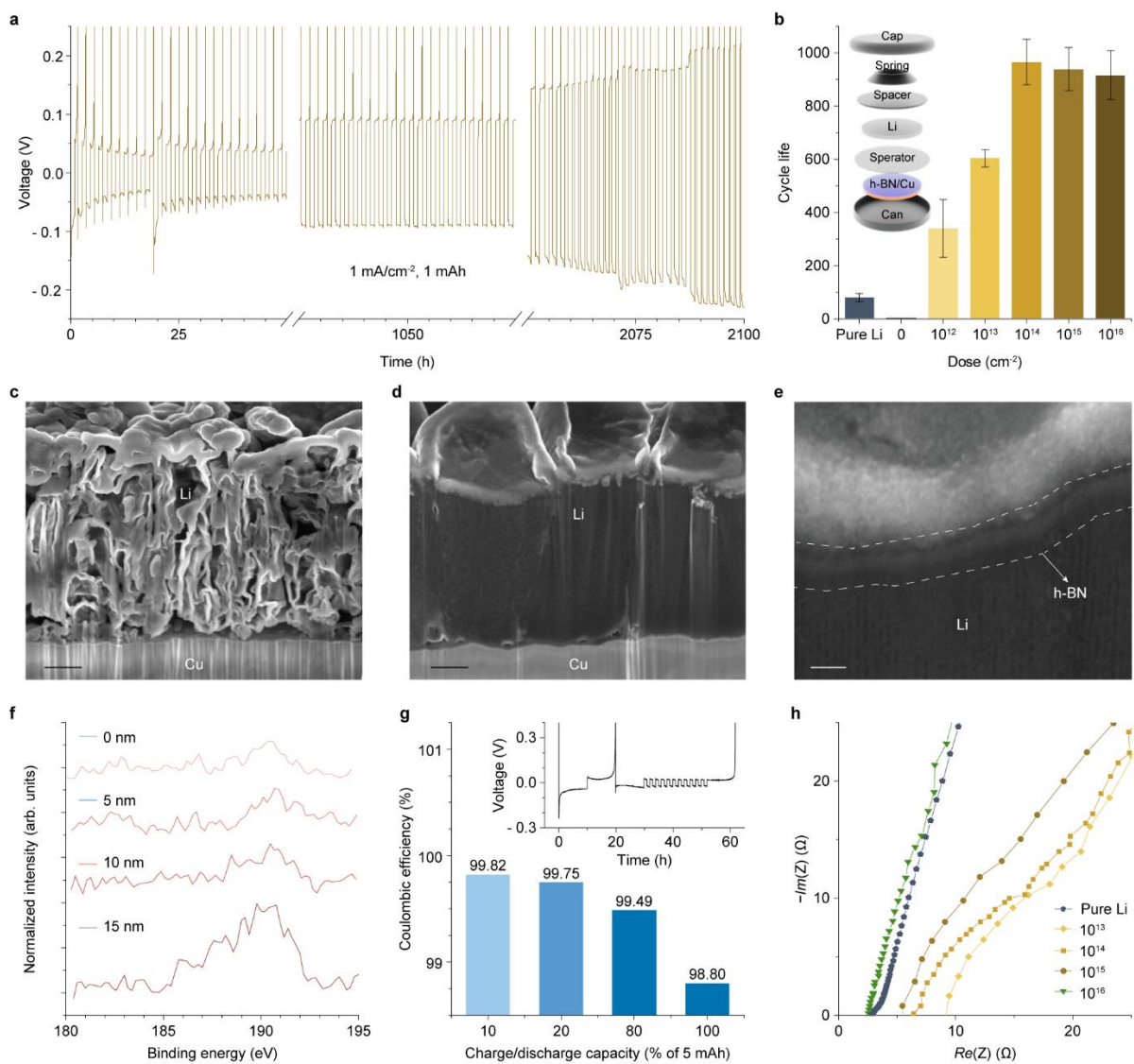

**Figure 4. Electrochemical characterization of irradiated h-BN protected lithium anode.**
**a**, Galvanostatic cycling voltage profile of lithium metal anode and copper substrates coated with CVD-grown h-BN irradiated at a dose of $10^{14}$ cm$^{-2}$ Ar$^+$ ions. Cycling was performed at a current density of 1 mA cm$^{-2}$ with a fixed capacity of 1 mAh. **b**, Statistical analysis of cycle lifetimes, defined as the number of cycles at which capacity retention falls to 80% of the initial value. Data represents at least three independent cells per condition, with error bars indicating standard deviation. **c–d**, Cross-sectional SEM images of lithium deposited on bare copper (**c**) and copper substrates protected by irradiated h-BN (**d**). Scale bars: 1 μm. **e**, High-magnification SEM image showing lithium plated beneath the irradiated h-BN layer (highlighted by white dashed lines). Scale bar: 200 nm. For panels c–d, the stage was tilted during imaging; thus, the actual size differs from the dimensions shown in the figures. **f**, X-ray photoelectron spectroscopy (XPS) analysis of the B 1s peak during progressive etching of the lithium-plated, irradiated h-BN electrode. **g**, Coulombic efficiency measured at different charge capacities using Aurbach's method of a sample with $10^{14}$ cm$^{-2}$ Ar$^+$ irradiation. Insets: voltage profile for CE measurement. **h**, EIS results for electrodes featuring h-BN irradiated at various doses. In the high frequency limit, Re (Z) between 5 − 10 Ω using a test current of 1 mamp corresponds to a very low voltage overpotential of 5 – 10 mV.

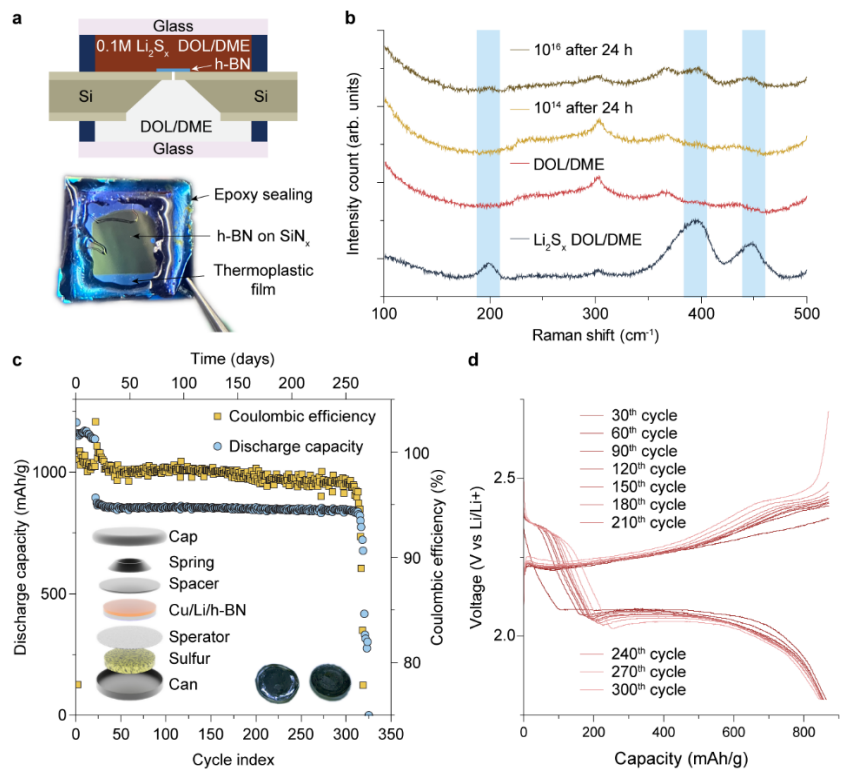

**Figure 5. Characterization of Li–S full cells employing irradiation-engineered h-BN.**
**a**, Schematic illustration and photograph of the experimental setup used to evaluate lithium polysulfide diffusion. The device consists of an h-BN flake sandwiched between two electrolyte reservoirs. **b**, Raman spectra obtained after 24 hours from: (i) lithium polysulfide dissolved in DOL/DME electrolyte, (ii) pure DOL/DME electrolyte, and (iii) electrolyte chambers separated by h-BN flakes irradiated at doses of $10^{14}$ and $10^{16}$ ions/cm$^2$. **c**, Galvanostatic cycling performance of Li-S batteries assembled with irradiation-engineered h-BN ($10^{14}$ ions/cm$^2$) as the Li-anode protective layer. Cells underwent 20 initial activation cycles at 1160 mAh/g (0.1 C), followed by long-term cycling at 870 mAh/g (0.1 C). The current density was maintained at 0.8 mA/cm$^2$. Inset: photograph of the disassembled battery after reaching end-of-life due to electrolyte depletion. **d**, Voltage profiles recorded across various cycle indices.

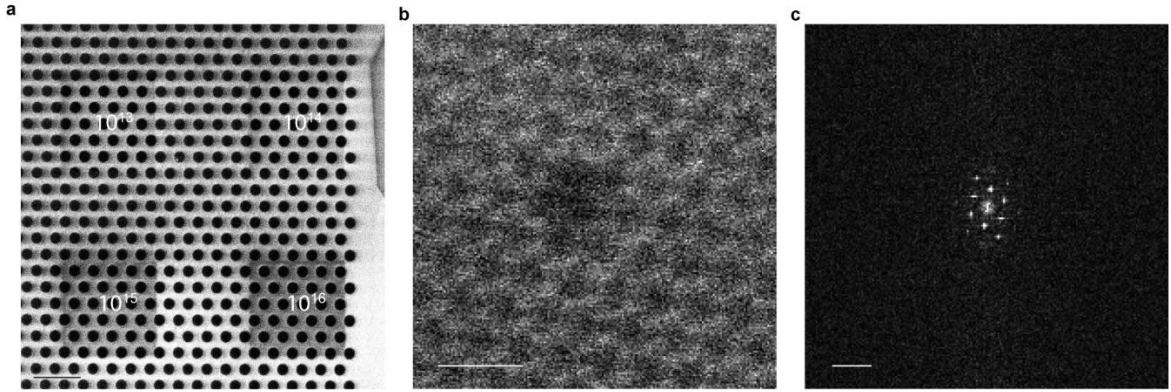

**Extended Data Fig. 1. STEM images of the defects in h-BN.**
**a**, The SEM image of patterned ion irradiation of a h-BN flake on holey silicon nitride grid. The irradiation doses (in cm$^{-2}$) are labeled in the image. Scale bar: 5 µm. **b,** The raw data of the defect in Fig. 2c. Scale bar: 500 pm. **c,** Fourier transform of panel **a**. Scale bar: 10 1/nm.

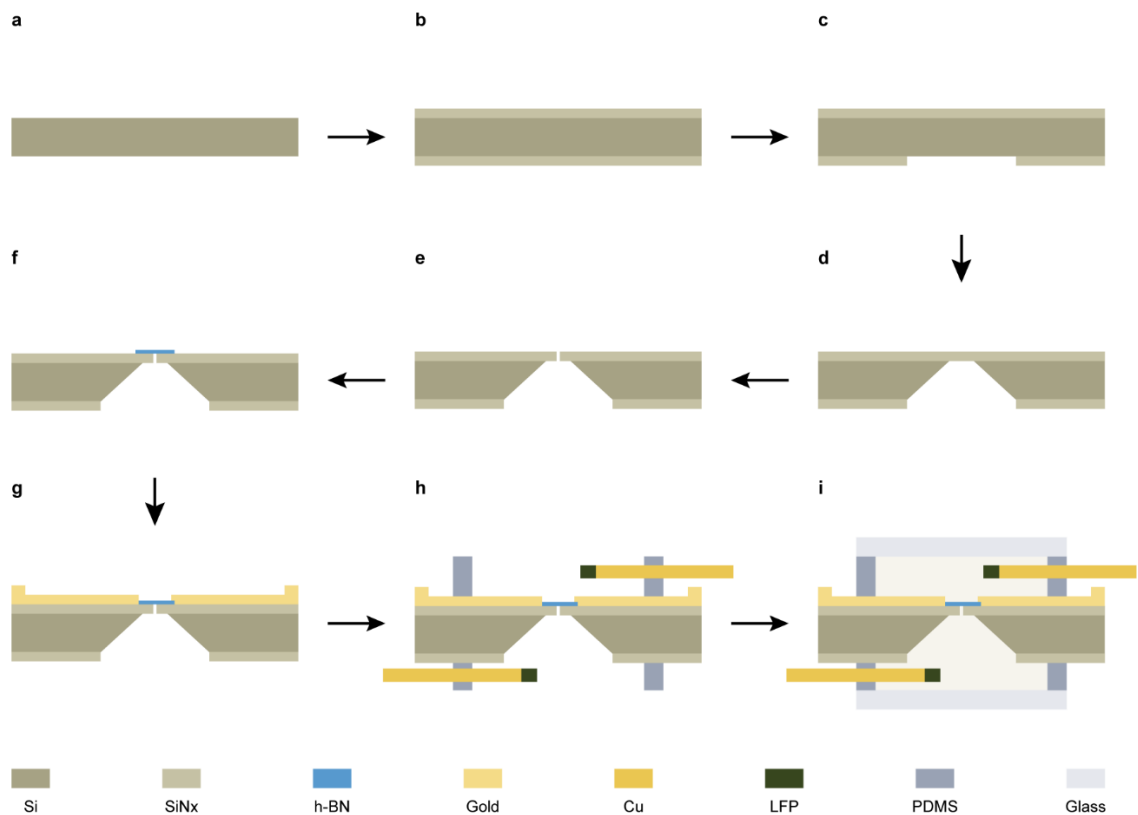

**Extended Data Fig. 2. Fabrication processes.**
**a**, Start with a double-side polished silicon substrate. **b**, Grow ~ 250 nm of silicon nitride on both sides of the substrate. **c**, Perform photolithography and dry etching to create a ~600 μm-wide window in the silicon nitride layer on the substrate's backside. **d**, Wet-etch the silicon exposed in the window using 30% KOH solution. **e**, Employ photolithography and dry etching to open a 10 μm-diameter hole on the front-side silicon nitride layer. **f**, Deterministically transfer an exfoliated h-BN flake onto the hole, then perform plasma-focused ion beam (PFIB) irradiation at the desired dose. **g**, Fabricate metal contacts on the h-BN using photolithography and metal deposition. **h**, Form electrolyte reservoirs by constructing PDMS sidewalls, inserting lithium iron phosphate (LFP)-coated copper foil electrodes. **i**, Inject electrolyte into the reservoirs and seal them with glass coverslips.

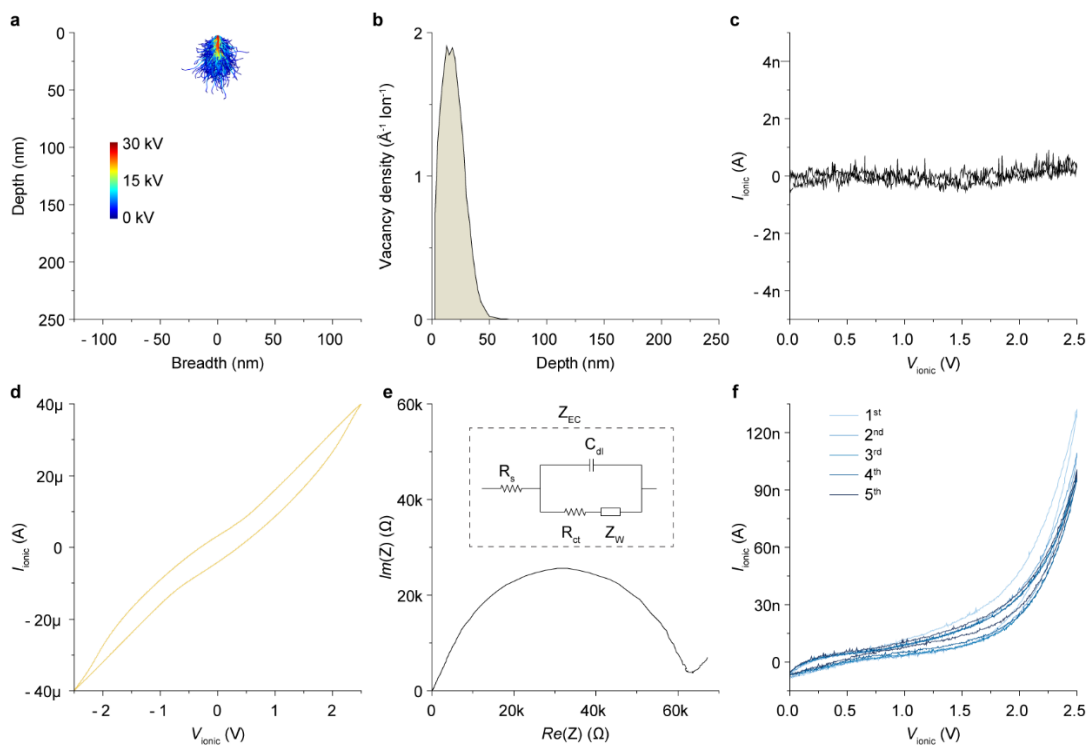

**Extended Data Fig 3. Irradiation of silicon nitride substrate.**

**a**, Simulated trajectories of 1,000 $Ar^+$ ions accelerated at 30 kV, impinging at normal incidence onto a 250 × 250 nm $Si_3N_4$ film. Ion trajectories are color-coded by energy. **b**, Simulated vacancy-density profile across the depth of the $Si_3N_4$ layer resulting from 30 kV $Ar^+$ irradiation. **c**, Ionic current measured across the $Si_3N_4$ film irradiated at a dose of $10^{16}$ $cm^{-2}$, demonstrating the film's intrinsic ionic insulation. **d,** Cyclic voltammetry of a cell containing an open hole in the silicon nitride window (without h-BN) at a fast scan rate (100 mV/s)fast enough to suppress the faradaic reactions and emphasize the resistance of the ion/charge transport. **e**, Impedance spectrum of a reference cell containing an open hole in the silicon nitride window (without h-BN). For the equivalent circuit, $R_s$ is series resistor, $R_{ct}$ is the charge transfer resistor, $C_{dl}$ is the double layer capacitor, and $Z_W$ is the Warburg impedance. **f,** Ionic current measured over five consecutive cycles for an h-BN flake irradiated at a dose of $10^{15}$ $cm^{-2}$. The current decreases after the initial cycle and subsequently stabilizes.

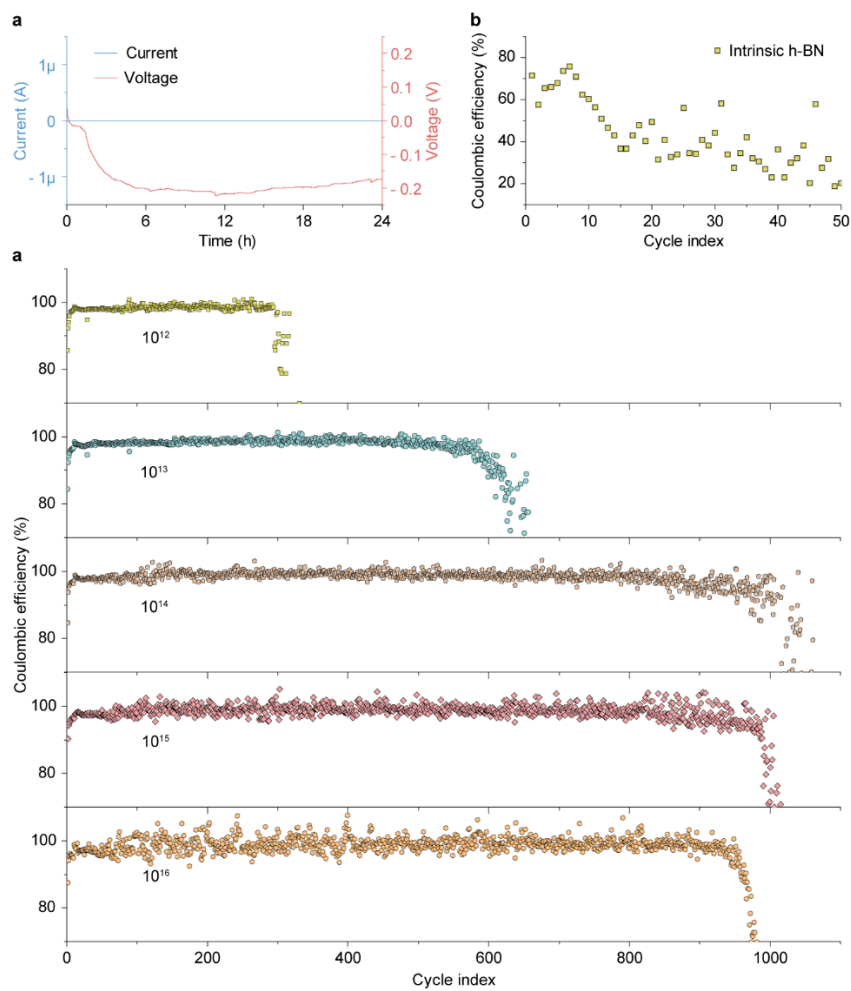

**Extended Data Fig. 4. Electrochemical performance of half-cell batteries with h-BN.**
**a**, Voltage profile of a half-cell battery assembled with a 30–40 nm-thick intrinsic h-BN layer under a constant current of 1 mA. Due to the intrinsic ionic insulation of pristine h-BN, no effective current could pass, resulting in a floating voltage. **b**, Galvanostatic cycling of a cell with approximately 10 nm-thick intrinsic h-BN. The presence of grain boundaries and defects in this thinner film allowed limited ion conduction, enabling cycling but resulting in significantly reduced battery performance. **c,** Galvanostatic cycling performance of half-cells comprising lithium metal paired with copper substrates with CVD-grown h-BN irradiated at various $Ar^+$ ion doses. Cycling was conducted at a current density of 1 mA cm$^{-2}$ and a fixed capacity of 1 mAh.

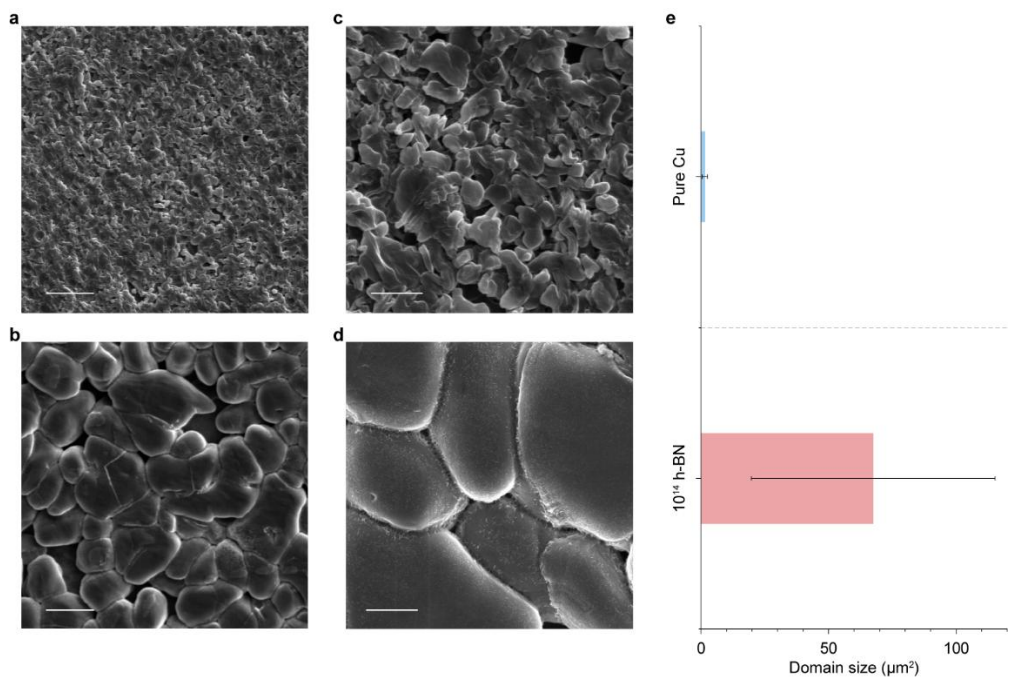

**Extended Data Fig. 5. SEM characterizations of plated lithium.**
**a-b**, Top view of the plated lithium on bare copper (a) and copper with irradiated h-BN (b). Scale bars: 10 um.
**c-d**, High magnification top view of the plated lithium on bare copper (c) and copper with irradiated h-BN (d). Scale bars: 3 um. **e,** Statistical representation of the domain area size. Error bars indicate the standard deviation based on measurements from over 20 domains per condition.

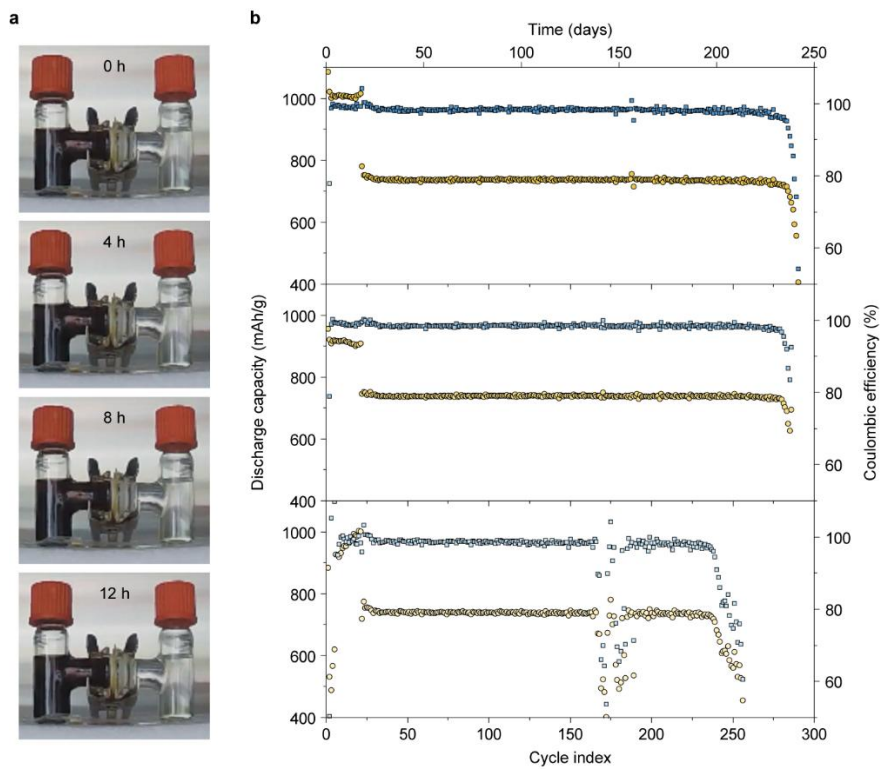

**Extended Data Fig.6. Additional Li-S cell performance.**
a, Photographs of H-type glass cells assembled by the PP/PE/PP separator with irradiated h-BN (thickness: 35nm) with a dose of $10^{14}$ ions/cm$^2$. b, Three additional Li-S coin cells utilizing h-BN with an irradiation dose of $10^{14}$ cm$^{-2}$ were tested at a charge and discharge rate of 0.1 C. The third cell (bottom panel) exhibited issues starting from the first cycle, possibly due to assembly or materials issues, and was therefore excluded from statistical analysis as an outlier.